\magnification=\magstephalf
\newbox\SlashedBox 
\def\slashed#1{\setbox\SlashedBox=\hbox{#1}
\hbox to 0pt{\hbox to 1\wd\SlashedBox{\hfil/\hfil}\hss}{#1}}
\def\hboxtosizeof#1#2{\setbox\SlashedBox=\hbox{#1}
\hbox to 1\wd\SlashedBox{#2}}

\def\dfrac#1/#2{%
\hskip-.2em\kern .2em\raise.4ex\hbox{\the\scriptfont0 #1}\kern-.2em/%
\kern -.15em\lower.35ex\hbox{\the\scriptfont0 #2}}
\def\mathslashed#1{\setbox\SlashedBox=\hbox{$#1$}
\hbox to 0pt{\hbox to 1\wd\SlashedBox{\hfil/\hfil}\hss}#1}

\def\clap#1{\hbox to 0pt{\hss#1\hss}}

\def\ifsmall{\iffalse}  
\def\titlepagefont{}  

\def\DefineTeXgraphics{%
\special{ps::[global] /TeXgraphics { } def}}  

\def\today{\ifcase\month\or January\or February\or March\or April\or May
\or June\or July\or August\or September\or October\or November\or
December\fi\space\number\day, \number\year}
\def\eatPrefix19{}
\def\Year{\expandafter\eatPrefix\the\year}
\newcount\hours \newcount\minutes
\def\monthname{\ifcase\month\or
January\or February\or March\or April\or May\or June\or July\or
August\or September\or October\or November\or December\fi}
\def\shortmonthname{\ifcase\month\or
Jan\or Feb\or Mar\or Apr\or May\or Jun\or Jul\or
Aug\or Sep\or Oct\or Nov\or Dec\fi}

\def\TimeStamp{\hours\the\time\divide\hours by60%
\minutes -\the\time\divide\minutes by60\multiply\minutes by60%
\advance\minutes by\the\time%
${\rm \shortmonthname}\cdot\if\day<10{}0\fi\the\day\cdot\the\year%
\qquad\the\hours:\if\minutes<10{}0\fi\the\minutes$}




\def\Title#1{%
\vskip 1in{\titlefont\centerline{#1}}\vskip .5in}
 
\def\Date#1{\leftline{#1}\tenrm\supereject%
\global\hsize=\hsbody\global\hoffset=\hbodyoffset%
\footline={\hss\tenrm\folio\hss}}

\newif\ifdraftmode
\newif\ifleftlabels  

\def\nolabels{\def\wrlabeL##1{}\def\eqlabeL##1{}\def\reflabeL##1{}}
\def\writelabels{\def\wrlabeL##1{\leavevmode\vadjust{\rlap{\smash%
{\line{{\escapechar=` \hfill\rlap{\sevenrm\hskip.03in\string##1}}}}}}}%
\def\eqlabeL##1{{\escapechar-1\rlap{\sevenrm\hskip.05in\string##1}}}%
\def\reflabeL##1{\noexpand\rlap{\noexpand\sevenrm[\string##1]}}}
\def\writeleftlabels{\def\wrlabeL##1{\leavevmode\vadjust{\rlap{\smash%
{\line{{\escapechar=` \hfill\rlap{\sevenrm\hskip.03in\string##1}}}}}}}%
\def\eqlabeL##1{{\escapechar-1%
\rlap{\sixrm\hskip.05in\string##1}%
\llap{\sevenrm\string##1\hskip.03in\hbox to \hsize{}}}}%
\def\reflabeL##1{\noexpand\rlap{\noexpand\sevenrm[\string##1]}}}
\nolabels

\input hyperbasics.tex

\newdimen\fullhsize
\newdimen\hstitle
\hstitle=\hsize 
\newdimen\hsbody
\hsbody=\hsize 
\newdimen\hbodyoffset
\hbodyoffset=\hoffset 
\newbox\leftpage
\def\abstract#1{#1}
\def\rotated{\special{ps: landscape}
\magnification=1000  
\baselineskip=14pt
\global\hstitle=9truein\global\hsbody=4.75truein
\global\vsize=7truein\global\voffset=-.31truein
\global\hoffset=-0.54in\global\hbodyoffset=-.54truein
\global\fullhsize=10truein
\def\DefineTeXgraphics{%
\special{ps::[global] 
/TeXgraphics {currentpoint translate 0.7 0.7 scale
              -80 0.72 mul -1000 0.72 mul translate} def}}
\let\lr=L
\def\ifsmall{\iftrue}
\def\titlepagefont{\twelvepoint}
\trueseventeenpoint
\def\almostshipout##1{\if L\lr \count1=1
      \global\setbox\leftpage=##1 \global\let\lr=R
   \else \count1=2
      \shipout\vbox{\hbox to\fullhsize{\box\leftpage\hfil##1}}
      \global\let\lr=L\fi}

\output={\ifnum\count0=1 
 \shipout\vbox{\hbox to \fullhsize{\hfill\pagebody\hfill}}\advancepageno
 \else
 \almostshipout{\leftline{\vbox{\pagebody\makefootline}}}\advancepageno 
 \fi}

\def\abstract##1{{\leftskip=1.5in\rightskip=1.5in ##1\par}} }

\def\linemessage#1{\immediate\write16{#1}}

\global\newcount\secno \global\secno=0
\global\newcount\appno \global\appno=0
\global\newcount\meqno \global\meqno=1
\global\newcount\subsecno \global\subsecno=0
\global\newcount\figno \global\figno=0

\newif\ifAnyCounterChanged
\let\terminator=\relax
\def\normalize#1{\ifx#1\terminator\let\next=\relax\else%
\if#1i\aftergroup i\else\if#1v\aftergroup v\else\if#1x\aftergroup x%
\else\if#1l\aftergroup l\else\if#1c\aftergroup c\else%
\if#1m\aftergroup m\else%
\if#1I\aftergroup I\else\if#1V\aftergroup V\else\if#1X\aftergroup X%
\else\if#1L\aftergroup L\else\if#1C\aftergroup C\else%
\if#1M\aftergroup M\else\aftergroup#1\fi\fi\fi\fi\fi\fi\fi\fi\fi\fi\fi\fi%
\let\next=\normalize\fi%
\next}
\def\makeNormal#1#2{\def\doNormalDef{\edef#1}\begingroup%
\aftergroup\doNormalDef\aftergroup{\normalize#2\terminator\aftergroup}%
\endgroup}

\def\warnIfChanged#1#2{%
\ifundef#1
\else\begingroup%
\edef\oldDefinitionOfCounter{#1}\edef\newDefinitionOfCounter{#2}%
\ifx\oldDefinitionOfCounter\newDefinitionOfCounter%
\else%
\linemessage{Warning: definition of \noexpand#1 has changed.}%
\global\AnyCounterChangedtrue\fi\endgroup\fi}

\def\Section#1{\global\advance\secno by1\relax\global\meqno=1%
\global\subsecno=0%
\bigbreak\bigskip
\centerline{\twelvepoint \bf %
\the\secno. #1}%
\par\nobreak\medskip\nobreak}
\def\tagsection#1{%
\warnIfChanged#1{\the\secno}%
\xdef#1{\the\secno}%
\ifWritingAuxFile\immediate\write\auxfile{\noexpand\xdef\noexpand#1{#1}}\fi%
}
\def\section{\Section}
\def\Subsection#1{\global\advance\subsecno by1\relax\medskip %
\leftline{\bf\the\secno.\the\subsecno\ #1}%
\par\nobreak\smallskip\nobreak}
\def\tagsubsection#1{%
\warnIfChanged#1{\the\secno.\the\subsecno}%
\xdef#1{\the\secno.\the\subsecno}%
\ifWritingAuxFile\immediate\write\auxfile{\noexpand\xdef\noexpand#1{#1}}\fi%
}

\def\subsection{\Subsection}

\def\romappno{\uppercase\expandafter{\romannumeral\appno}}
\def\makeNormalizedRomappno{%
\expandafter\makeNormal\expandafter\normalizedromappno%
\expandafter{\romannumeral\appno}%
\edef\normalizedromappno{\uppercase{\normalizedromappno}}}
\def\Appendix#1{\global\advance\appno by1\relax\global\meqno=1\global\secno=0%
\global\subsecno=0%
\bigbreak\bigskip
\centerline{\twelvepoint \bf Appendix %
\romappno. #1}%
\par\nobreak\medskip\nobreak}
\def\tagappendix#1{\makeNormalizedRomappno%
\warnIfChanged#1{\normalizedromappno}%
\xdef#1{\normalizedromappno}%
\ifWritingAuxFile\immediate\write\auxfile{\noexpand\xdef\noexpand#1{#1}}\fi%
}
\def\appendix{\Appendix}
\def\Subappendix#1{\global\advance\subsecno by1\relax\medskip %
\leftline{\bf\romappno.\the\subsecno\ #1}%
\par\nobreak\smallskip\nobreak}
\def\tagsubappendix#1{\makeNormalizedRomappno%
\warnIfChanged#1{\normalizedromappno.\the\subsecno}%
\xdef#1{\normalizedromappno.\the\subsecno}%
\ifWritingAuxFile\immediate\write\auxfile{\noexpand\xdef\noexpand#1{#1}}\fi%
}

\def\eqn#1{\makeNormalizedRomappno%
\ifnum\secno>0%
  \warnIfChanged#1{\the\secno.\the\meqno}%
  \eqno(\the\secno.\the\meqno)\xdef#1{\the\secno.\the\meqno}%
     \global\advance\meqno by1
\else\ifnum\appno>0%
  \warnIfChanged#1{\normalizedromappno.\the\meqno}%
  \eqno({\rm\romappno}.\the\meqno)%
      \xdef#1{\normalizedromappno.\the\meqno}%
     \global\advance\meqno by1
\else%
  \warnIfChanged#1{\the\meqno}%
  \eqno(\the\meqno)\xdef#1{\the\meqno}%
     \global\advance\meqno by1
\fi\fi%
\eqlabeL#1%
\ifWritingAuxFile\immediate\write\auxfile{\noexpand\xdef\noexpand#1{#1}}\fi%
}
\def\defeqn#1{\makeNormalizedRomappno%
\ifnum\secno>0%
  \warnIfChanged#1{\the\secno.\the\meqno}%
  \xdef#1{\the\secno.\the\meqno}%
     \global\advance\meqno by1
\else\ifnum\appno>0%
  \warnIfChanged#1{\normalizedromappno.\the\meqno}%
  \xdef#1{\normalizedromappno.\the\meqno}%
     \global\advance\meqno by1
\else%
  \warnIfChanged#1{\the\meqno}%
  \xdef#1{\the\meqno}%
     \global\advance\meqno by1
\fi\fi%
\eqlabeL#1%
\ifWritingAuxFile\immediate\write\auxfile{\noexpand\xdef\noexpand#1{#1}}\fi%
}
\def\anoneqn{\makeNormalizedRomappno%
\ifnum\secno>0
  \eqno(\the\secno.\the\meqno)%
     \global\advance\meqno by1
\else\ifnum\appno>0
  \eqno({\rm\normalizedromappno}.\the\meqno)%
     \global\advance\meqno by1
\else
  \eqno(\the\meqno)%
     \global\advance\meqno by1
\fi\fi%
}
\def\mfig#1#2{\ifx#20
\else\global\advance\figno by1%
\relax#1\the\figno%
\warnIfChanged#2{\the\figno}%
\xdef#2{\the\figno}%
\reflabeL#2%
\ifWritingAuxFile\immediate\write\auxfile{\noexpand\xdef\noexpand#2{#2}}\fi\fi%
}

\catcode`@=11 

\newif\ifFiguresInText\FiguresInTexttrue
\newif\if@FigureFileCreated
\newwrite\capfile
\newwrite\figfile

\newif\ifcaption
\captiontrue
\def\captionsize{\tenrm}
\def\PlaceTextFigure#1#2#3#4{%
\vskip 0.5truein%
\noindent#3\hfil\epsfbox{#4}\hfil\break%
\ifcaption\vskip 5pt\noindent\hfil\vbox{\captionsize \noindent Figure #1. #2}\hfil\fi%
\vskip10pt}
\def\PlaceEndFigure#1#2{%
\epsfxsize=\hsize\epsfbox{#2}\vfill\centerline{Figure #1.}\eject}

\def\LoadFigure#1#2#3#4{%
\vphantom{\mfig{}#1}
\ifx#10
\else
\fi
\ifFiguresInText
\PlaceTextFigure{#1}{#2}{#3}{#4}%
\else
\if@FigureFileCreated\else%
\immediate\openout\capfile=\jobname.caps%
\immediate\openout\figfile=\jobname.figs%
@FigureFileCreatedtrue\fi%
\immediate\write\capfile{\noexpand\item{Figure \noexpand#1.\ }{#2}\vskip10pt}%
\immed	iate\write\figfile{\noexpand\PlaceEndFigure\noexpand#1{\noexpand#4}}%
\fi}

\def\listfigs{\ifFiguresInText\else%
\vfill\eject\immediate\closeout\capfile
\immediate\closeout\figfile%
\centerline{{\bf Figures}}\bigskip\frenchspacing%
\catcode`@=11 
\def\captionsize{\tenrm}
\input \jobname.caps\vfill\eject\nonfrenchspacing%
\catcode`\@=\active
\catcode`@=12  
\input\jobname.figs\fi}

\font\ninerm=cmr9
\font\eightrm=cmr8
\font\sixrm=cmr6

\def\loadtrueseventeenpoint{
 \font\seventeenrm=cmr10 at 17.28truept
 \font\seventeeni=cmmi10 at 17.28truept
 \font\seventeenbf=cmbx10 at 17.28truept
 \font\seventeenit=cmti10 at 17.28truept
 \font\seventeensl=cmsl10 at 17.28truept
 \font\seventeensy=cmsy10 at 17.28truept
}
\def\loadfourteenpoint{
\font\fourteenrm=cmr10 at 14.4pt
\font\fourteeni=cmmi10 at 14.4pt
\font\fourteenit=cmti10 at 14.4pt
\font\fourteensl=cmsl10 at 14.4pt
\font\fourteensy=cmsy10 at 14.4pt
\font\fourteenbf=cmbx10 at 14.4pt
}
\def\loadtruetwelvepoint{
\font\twelverm=cmr10 at 12truept
\font\twelvei=cmmi10 at 12truept
\font\twelveit=cmti10 at 12truept
\font\twelvesl=cmsl10 at 12truept
\font\twelvesy=cmsy10 at 12truept
\font\twelvebf=cmbx10 at 12truept
\font\twelvesc=cmcsc10 at 12truept
}

\font\ninei=cmmi9
\font\eighti=cmmi8
\font\sixi=cmmi6
\skewchar\ninei='177 \skewchar\eighti='177 \skewchar\sixi='177

\font\ninesy=cmsy9
\font\eightsy=cmsy8
\font\sixsy=cmsy6
\skewchar\ninesy='60 \skewchar\eightsy='60 \skewchar\sixsy='60

\font\ninebf=cmbx9
\font\eightbf=cmbx8
\font\sixbf=cmbx6

\font\ninett=cmtt9
\font\eighttt=cmtt8

\hyphenchar\tentt=-1 
\hyphenchar\ninett=-1
\hyphenchar\eighttt=-1         

\font\ninesl=cmsl9
\font\eightsl=cmsl8

\font\nineit=cmti9
\font\eightit=cmti8
\font\sevenit=cmti7

\scriptfont\itfam=\sevenit


                      
\newskip\ttglue
\def\tenpoint{\def\rm{\fam0\tenrm}%
  \textfont0=\tenrm \scriptfont0=\sevenrm \scriptscriptfont0=\fiverm
  \textfont1=\teni \scriptfont1=\seveni \scriptscriptfont1=\fivei
  \textfont2=\tensy \scriptfont2=\sevensy \scriptscriptfont2=\fivesy
  \textfont3=\tenex \scriptfont3=\tenex \scriptscriptfont3=\tenex
  \def\it{\fam\itfam\tenit}%
      \textfont\itfam=\tenit\scriptfont\itfam=\sevenit
  \def\sl{\fam\slfam\tensl}\textfont\slfam=\tensl
  \def\bf{\fam\bffam\tenbf}\textfont\bffam=\tenbf \scriptfont\bffam=\sevenbf
  \scriptscriptfont\bffam=\fivebf
  \normalbaselineskip=12pt
  \let\sc=\eightrm
  \let\big=\tenbig
  \setbox\strutbox=\hbox{\vrule height8.5pt depth3.5pt width\z@}%
  \normalbaselines\rm}

\def\twelvepoint{\def\rm{\fam0\twelverm}%
  \textfont0=\twelverm \scriptfont0=\ninerm \scriptscriptfont0=\sevenrm
  \textfont1=\twelvei \scriptfont1=\ninei \scriptscriptfont1=\seveni
  \textfont2=\twelvesy \scriptfont2=\ninesy \scriptscriptfont2=\sevensy
  \textfont3=\tenex \scriptfont3=\tenex \scriptscriptfont3=\tenex
  \def\it{\fam\itfam\twelveit}\textfont\itfam=\twelveit
  \def\sl{\fam\slfam\twelvesl}\textfont\slfam=\twelvesl
  \def\bf{\fam\bffam\twelvebf}\textfont\bffam=\twelvebf%
  \scriptfont\bffam=\ninebf
  \scriptscriptfont\bffam=\sevenbf
  \normalbaselineskip=12pt
  \let\sc=\eightrm
  \let\big=\tenbig
  \setbox\strutbox=\hbox{\vrule height8.5pt depth3.5pt width\z@}%
  \normalbaselines\rm}

\def\fourteenpoint{\def\rm{\fam0\fourteenrm}%
  \textfont0=\fourteenrm \scriptfont0=\tenrm \scriptscriptfont0=\sevenrm
  \textfont1=\fourteeni \scriptfont1=\teni \scriptscriptfont1=\seveni
  \textfont2=\fourteensy \scriptfont2=\tensy \scriptscriptfont2=\sevensy
  \textfont3=\tenex \scriptfont3=\tenex \scriptscriptfont3=\tenex
  \def\it{\fam\itfam\fourteenit}\textfont\itfam=\fourteenit
  \def\sl{\fam\slfam\fourteensl}\textfont\slfam=\fourteensl
  \def\bf{\fam\bffam\fourteenbf}\textfont\bffam=\fourteenbf%
  \scriptfont\bffam=\tenbf
  \scriptscriptfont\bffam=\sevenbf
  \normalbaselineskip=17pt
  \let\sc=\elevenrm
  \let\big=\tenbig                                          
  \setbox\strutbox=\hbox{\vrule height8.5pt depth3.5pt width\z@}%
  \normalbaselines\rm}

\def\seventeenpoint{\def\rm{\fam0\seventeenrm}%
  \textfont0=\seventeenrm \scriptfont0=\fourteenrm \scriptscriptfont0=\tenrm
  \textfont1=\seventeeni \scriptfont1=\fourteeni \scriptscriptfont1=\teni
  \textfont2=\seventeensy \scriptfont2=\fourteensy \scriptscriptfont2=\tensy
  \textfont3=\tenex \scriptfont3=\tenex \scriptscriptfont3=\tenex
  \def\it{\fam\itfam\seventeenit}\textfont\itfam=\seventeenit
  \def\sl{\fam\slfam\seventeensl}\textfont\slfam=\seventeensl
  \def\bf{\fam\bffam\seventeenbf}\textfont\bffam=\seventeenbf%
  \scriptfont\bffam=\fourteenbf
  \scriptscriptfont\bffam=\twelvebf
  \normalbaselineskip=21pt
  \let\sc=\fourteenrm
  \let\big=\tenbig                                          
  \setbox\strutbox=\hbox{\vrule height 12pt depth 6pt width\z@}%
  \normalbaselines\rm}

\def\ninepoint{\def\rm{\fam0\ninerm}%
  \textfont0=\ninerm \scriptfont0=\sixrm \scriptscriptfont0=\fiverm
  \textfont1=\ninei \scriptfont1=\sixi \scriptscriptfont1=\fivei
  \textfont2=\ninesy \scriptfont2=\sixsy \scriptscriptfont2=\fivesy
  \textfont3=\tenex \scriptfont3=\tenex \scriptscriptfont3=\tenex
  \def\it{\fam\itfam\nineit}\textfont\itfam=\nineit
  \def\sl{\fam\slfam\ninesl}\textfont\slfam=\ninesl
  \def\bf{\fam\bffam\ninebf}\textfont\bffam=\ninebf \scriptfont\bffam=\sixbf
  \scriptscriptfont\bffam=\fivebf
  \normalbaselineskip=11pt
  \let\sc=\sevenrm
  \let\big=\ninebig
  \setbox\strutbox=\hbox{\vrule height8pt depth3pt width\z@}%
  \normalbaselines\rm}

\def\eightpoint{\def\rm{\fam0\eightrm}%
  \textfont0=\eightrm \scriptfont0=\sixrm \scriptscriptfont0=\fiverm%
  \textfont1=\eighti \scriptfont1=\sixi \scriptscriptfont1=\fivei%
  \textfont2=\eightsy \scriptfont2=\sixsy \scriptscriptfont2=\fivesy%
  \textfont3=\tenex \scriptfont3=\tenex \scriptscriptfont3=\tenex%
  \def\it{\fam\itfam\eightit}\textfont\itfam=\eightit%
  \def\sl{\fam\slfam\eightsl}\textfont\slfam=\eightsl%
  \def\bf{\fam\bffam\eightbf}\textfont\bffam=\eightbf \scriptfont\bffam=\sixbf%
  \scriptscriptfont\bffam=\fivebf%
  \normalbaselineskip=9pt%
  \let\sc=\sixrm%
  \let\big=\eightbig%
  \setbox\strutbox=\hbox{\vrule height7pt depth2pt width\z@}%
  \normalbaselines\rm}
  \let\sc=\eightrm

\def\tenbig#1{{\hbox{$\left#1\vbox to8.5pt{}\right.\n@space$}}}
\def\ninebig#1{{\hbox{$\textfont0=\tenrm\textfont2=\tensy
  \left#1\vbox to7.25pt{}\right.\n@space$}}}
\def\eightbig#1{{\hbox{$\textfont0=\ninerm\textfont2=\ninesy
  \left#1\vbox to6.5pt{}\right.\n@space$}}}

\def\footnote#1{\edef\@sf{\spacefactor\the\spacefactor}#1\@sf
      \insert\footins\bgroup\eightpoint
      \interlinepenalty100 \let\par=\endgraf
        \leftskip=\z@skip \rightskip=\z@skip
        \splittopskip=10pt plus 1pt minus 1pt \floatingpenalty=20000
        \smallskip\item{#1}\bgroup\strut\aftergroup\@foot\let\next}
\skip\footins=12pt plus 2pt minus 4pt 
\dimen\footins=30pc 

\newinsert\margin
\dimen\margin=\maxdimen
\def\titlefont{\seventeenpoint}
\loadtruetwelvepoint 
\loadtrueseventeenpoint

\def\eatOne#1{}
\def\ifundef#1{\expandafter\ifx%
\csname\expandafter\eatOne\string#1\endcsname\relax}
\def\notTrue{\iffalse}\def\isTrue{\iftrue}
\def\ifdef#1{{\ifundef#1%
\aftergroup\notTrue\else\aftergroup\isTrue\fi}}
\def\use#1{\ifundef#1\linemessage{Warning: \string#1 is undefined.}%
{\tt \string#1}\else#1\fi}



%
\catcode`"=11
\let\quote="
\catcode`"=12
\chardef\foo="22
\global\newcount\refno \global\refno=1
\newwrite\rfile
\newlinechar=`\^^J
\def\@ref#1#2{\the\refno\n@ref#1{#2}}
\def\h@ref#1#2#3{\href{#3}{\the\refno}\n@ref#1{#2}}
\def\n@ref#1#2{\xdef#1{\the\refno}%
\ifnum\refno=1\immediate\openout\rfile=\jobname.refs\fi%
\immediate\write\rfile{\noexpand\item{[\noexpand#1]\ }#2.}%
\global\advance\refno by1}
\def\nref{\n@ref} 
\def\ref{\@ref}   
\def\hrref{\h@ref}
\def\lref#1#2{\the\refno\xdef#1{\the\refno}%
\ifnum\refno=1\immediate\openout\rfile=\jobname.refs\fi%
\immediate\write\rfile{\noexpand\item{[\noexpand#1]\ }#2\semi}%
\global\advance\refno by1}
\def\cref#1{\immediate\write\rfile{#1\semi}}

\def\preref#1#2{\gdef#1{\@ref#1{#2}}}

\def\semi{;\hfil\noexpand\break}

\def\listrefs{\vfill\eject\immediate\closeout\rfile
\centerline{{\bf References}}\bigskip\frenchspacing%
\input \jobname.refs\vfill\eject\nonfrenchspacing}

\def\inputAuxIfPresent#1{\immediate\openin1=#1
\ifeof1\message{No file \auxfileName; I'll create one.
}\else\closein1\relax\input\auxfileName\fi%
}
\def\NPB{Nucl.\ Phys.\ B}




\newif\ifWritingAuxFile
\newwrite\auxfile
\def\SetUpAuxFile{%
\xdef\auxfileName{\jobname.aux}%
\inputAuxIfPresent{\auxfileName}%
\WritingAuxFiletrue%
\immediate\openout\auxfile=\auxfileName}


\def\bye{\par\vfill\supereject%
\ifAnyCounterChanged\linemessage{
Some counters have changed.  Re-run tex to fix them up.}\fi%
\end}

\catcode`\@=\active
\catcode`@=12  
\catcode`\"=\active

\def\Tr{\mathop{\rm Tr}\nolimits}

\def\pol{\varepsilon}

\def\dl^#1_#2{\delta^{#1}{}_{#2}}

\def\mod{\mathrel{\rm mod}}

\catcode`@=11  
\def\meqalign#1{\,\vcenter{\openup1\jot\m@th
   \ialign{\strut\hfil$\displaystyle{##}$ && $\displaystyle{{}##}$\hfil
             \crcr#1\crcr}}\,}
\catcode`@=12  


\baselineskip 15pt
\overfullrule 0.5pt

\def\Tr{\mathop{\rm Tr}\nolimits}

\def\mod{\mathop{\rm mod}\nolimits}

\def\pol{\varepsilon}

\def\ksl{\slashed{k}}
\def\Ksl{\slashed{K}}

\def\spa#1.#2{\left\langle#1\,#2\right\rangle}
\def\spb#1.#2{\left[#1\,#2\right]}
\def\lor#1.#2{\left(#1\,#2\right)}
\def\sand#1.#2.#3{%
\left\langle\smash{#1}{\vphantom1}^{-}\right|{#2}%
\left|\smash{#3}{\vphantom1}^{-}\right\rangle}
\def\sandp#1.#2.#3{%
\left\langle\smash{#1}{\vphantom1}^{-}\right|{#2}%
\left|\smash{#3}{\vphantom1}^{+}\right\rangle}
\def\sandpp#1.#2.#3{%
\left\langle\smash{#1}{\vphantom1}^{+}\right|{#2}%
\left|\smash{#3}{\vphantom1}^{+}\right\rangle}
\def\sandpm#1.#2.#3{%
\left\langle\smash{#1}{\vphantom1}^{+}\right|{#2}%
\left|\smash{#3}{\vphantom1}^{-}\right\rangle}
\def\sandmp#1.#2.#3{%
   \left\langle\smash{#1}{\vphantom1}^{-}\right|{#2}%
    \left|\smash{#3}{\vphantom1}^{+}\right\rangle}
\catcode`@=11  
\def\meqalign#1{\,\vcenter{\openup1\jot\m@th
   \ialign{\strut\hfil$\displaystyle{##}$ && $\displaystyle{{}##}$\hfil
             \crcr#1\crcr}}\,}
\catcode`@=12  

\input epsf
\def\captionsize{\ninerm}

\SetUpAuxFile
\hfuzz 20pt
\overfullrule 0pt

\def\e{\epsilon}
\def\tree{{\rm tree\vphantom{p}}}
\def\dash{\hbox{-\kern-.02em}}

\def\dash{\hbox{-\kern-.02em}}

\def\llongrightarrow{%
\relbar\mskip-0.5mu\joinrel\mskip-0.5mu\relbar\mskip-0.5mu\joinrel\longrightarrow}
\def\inlimit^#1{\buildrel#1\over\llongrightarrow}
\def\frac#1#2{{#1\over #2}}

\preref\DixonTASI{L.\ Dixon, in 
{\it QCD \& Beyond: Proceedings of TASI '95}, 
ed. D.\ E.\ Soper (World Scientific, 1996) [hep-ph/9601359]}
\preref\Recurrence{F.\ A.\ Berends and W.\ T.\ Giele, 
Nucl.\ Phys.\ B306:759 (1988)}
\preref\HelicityRecurrence{
D.\ A.\ Kosower,
Nucl.\ Phys.\ B335:23 (1990)%
}
\preref\BernChalmers{
Z.\ Bern and G.\ Chalmers,
Nucl.\ Phys.\ B447:465 (1995) [hep-ph/9503236]%
}
\preref\ManganoParke{M.\ Mangano and S.\ J.\ Parke, Phys.\ Rep.\ 200:301 (1991)}
\preref\SingleAntenna{
D.\ A.\ Kosower,
Phys.\ Rev.\ D57:5410 (1998) [hep-ph/9710213]%
}
\preref\MultipleAntenna{
D.\ A.\ Kosower,
Phys.\ Rev.\ D67:116003 (2003) [hep-ph/0212097]%
}
\preref\SusyFour{Z. Bern, L. Dixon, D. C. Dunbar, and D. A. Kosower,
Nucl.\ Phys.\ B425:217 (1994) [hep-ph/9403226]}
\preref\AllOrdersCollinear{
D. A. Kosower,
Nucl.\ Phys.\ B552:319 (1999) [hep-ph/9901201]%
}
\preref\DDFM{
V.\ Del Duca, A.\ Frizzo and F.\ Maltoni,
Nucl.\ Phys.\ B568:211 (2000) [hep-ph/9909464]%
}
\preref\GloverCampbell{
J.\ M.\ Campbell and E.\ W.\ N.\ Glover,
Nucl.\ Phys.\ B527:264 (1998) [hep-ph/9710255]%
}
\preref\CataniGrazzini{
S.\ Catani and M.\ Grazzini,
Phys.\ Lett.\ B446:143 (1999) [hep-ph/9810389]\semi
S.\ Catani and M.\ Grazzini,
Nucl.\ Phys.\ B570:287 (2000) [hep-ph/9908523]%
}
\preref\BerendsGieleSoft{
F.\ A.\ Berends and W.\ T.\ Giele,
Nucl.\ Phys.\ B313:595 (1989)%
}
\preref\AltarelliParisi{G.\ Altarelli and G.\ Parisi, Nucl.\ Phys.\ B126:298 (1977)}
\preref\GieleGlover{W.\ T.\ Giele and E.\ W.\ N.\ Glover, 
Phys.\ Rev.\ D46:1980 (1992)}
\preref\GieleGloverKosower{
W.\ T.\ Giele, E.\ W.\ N.\ Glover and D.\ A.\ Kosower,
Nucl.\ Phys.\ B403:633 (1993) [hep-ph/9302225]%
}
\preref\CataniSeymour{
S.\ Catani and M.\ H.\ Seymour,
Phys.\ Lett.\ B378:287 (1996) [hep-ph/9602277]\semi
S.\ Catani and M.\ H.\ Seymour,
Nucl.\ Phys.\ B485:291 (1997); erratum-ibid.\ B510:503 (1997) [hep-ph/9605323]%
}
\preref\Color{%
F.\ A.\ Berends and W.\ T.\ Giele,
Nucl.\ Phys.\ B294:700 (1987)\semi
D.\ A.\ Kosower, B.-H.\ Lee and V.\ P.\ Nair, Phys.\ Lett.\ 201B:85 (1988)\semi
M.\ Mangano, S.\ Parke and Z.\ Xu, Nucl.\ Phys.\ B298:653 (1988)\semi
Z.\ Bern and D.\ A.\ Kosower, Nucl.\ Phys.\ B362:389 (1991)}
\preref\qqggg{Z. Bern, L. Dixon, and D. A. Kosower,
Nucl.\ Phys.\  B437:259 (1995) [hep-ph/9409393]}
\preref\AlternateColorDecomposition{
V.\ Del Duca, L.\ J.\ Dixon and F.\ Maltoni,
Nucl.\ Phys.\ B571: 51 (2000) [hep-ph/9910563]%
}

\preref\Spinor{%
F.\ A.\ Berends, R.\ Kleiss, P.\ De Causmaecker, R.\ Gastmans, and T.\ T.\ Wu,
        Phys.\ Lett.\ 103B:124 (1981)\semi
P.\ De Causmaeker, R.\ Gastmans,  W.\ Troost, and  T.\ T.\ Wu,
Nucl.\ Phys.\ B206:53 (1982)\semi
Z.\ Xu, D.-H.\ Zhang, L.\ Chang, Tsinghua University
                  preprint TUTP--84/3 (1984), unpublished\semi
R.\ Kleiss and W.\ J.\ Stirling, 
   Nucl.\ Phys.\ B262:235 (1985)\semi
   J.\ F.\ Gunion and Z.\ Kunszt, Phys.\ Lett.\ 161B:333 (1985)\semi
Z.\ Xu, D.-H.\ Zhang, and L.\ Chang, Nucl.\ Phys.\ B291:392 (1987)}

\preref\UnitarityAll{
Z.\ Bern, L.\ J.\ Dixon, D.\ C.\ Dunbar and D.\ A.\ Kosower,
Nucl.\ Phys.\ B435:59 (1995) [hep-ph/9409265].
}

\preref\UnitarityMachinery{
Z.\ Bern, L.\ J.\ Dixon, D.\ C.\ Dunbar and D.\ A.\ Kosower,
Nucl.\ Phys.\ B435:59 (1995) [hep-ph/9409265]\semi
Z.\ Bern, L.\ J.\ Dixon and D.\ A.\ Kosower,
Nucl.\ Phys.\ Proc.\ Suppl.\  51C:243 (1996)
[hep-ph/9606378]\semi
Z.\ Bern, L.\ J.\ Dixon and D.\ A.\ Kosower,
Ann.\ Rev.\ Nucl.\ Part.\ Sci.\  46:109 (1996)
[hep-ph/9602280]\semi
Z.\ Bern and A.\ G.\ Morgan,
Nucl.\ Phys.\ B467:479 (1996) [hep-ph/9511336]\semi%
Z.\ Bern, L.\ J.\ Dixon and D.\ A.\ Kosower,
hep-ph/0404293%
}

\preref\UnitarityB{%
Z.\ Bern, L.\ J.\ Dixon, D.\ C.\ Dunbar and D.\ A.\ Kosower,
Nucl.\ Phys.\ B435:59 (1995) [hep-ph/9409265].
}
\preref\UnitarityReview{
}

\preref\UnitarityCalculations{
Z.\ Bern, L.\ J.\ Dixon, D.\ C.\ Dunbar and D.\ A.\ Kosower,
Nucl.\ Phys.\ B{425}:217 (1994)
[hep-ph/9403226]\semi
Z.\ Bern, L.\ J.\ Dixon and D.\ A.\ Kosower,
Nucl.\ Phys.\ B{437}:259 (1995)
[hep-ph/9409393]\semi
Z.\ Bern, L.\ J.\ Dixon, D.\ A.\ Kosower and S.\ Weinzierl,
Nucl.\ Phys.\ B{489}:3 (1997)
[hep-ph/9610370]\semi
Z.\ Bern, L.\ J.\ Dixon and D.\ A.\ Kosower,
Nucl.\ Phys.\ B{513}:3 (1998)
[hep-ph/9708239]\semi
Z.\ Bern, L.\ J.\ Dixon and D.\ A.\ Kosower,
JHEP {0001}:027 (2000)
[hep-ph/0001001]\semi
Z.\ Bern, A.\ De Freitas and L.\ J.\ Dixon,
JHEP {0109}:037 (2001)
[hep-ph/0109078]\semi
Z.\ Bern, A.\ De Freitas and L.\ J.\ Dixon,
JHEP {0203}:018 (2002)
[hep-ph/0201161]\semi
Z.\ Bern, A.\ De Freitas and L.\ J.\ Dixon,
JHEP {0306}:028 (2003)
[hep-ph/0304168]\semi
C.\ Anastasiou, Z.\ Bern, L.\ J.\ Dixon and D.\ A.\ Kosower,
Phys.\ Rev.\ Lett.\  91:251602 (2003)
[hep-th/0309040]%
}

\preref\HigherLoopAntenna{
D.\ A.\ Kosower,
hep-ph/0301069%
}
\preref\BernMorgan{
Z.\ Bern and A.\ G.\ Morgan,
Nucl.\ Phys.\ B467:479 (1996) [hep-ph/9511336]%
}

\preref\CataniConjecture{
S.\ Catani,
Phys.\ Lett.\ B427:161 (1998) [hep-ph/9802439]%
}
\preref\StermanTejeda{
G.\ Sterman and M.\ E.\ Tejeda-Yeomans,
Phys.\ Lett.\ B552:48 (2003) [hep-ph/0210130]%
}
\preref\vanNeerven{W.\ L.\ van\ Neerven, \NPB 268:453 (1986)}
\preref\CataniGrazziniSoft{
S.\ Catani and M.\ Grazzini,
Nucl.\ Phys.\ B591:435 (2000) [hep-ph/0007142]%
}
\preref\FKS{
S.\ Frixione, Z.\ Kunszt and A.\ Signer,
Nucl.\ Phys.\ B467:399 (1996) [hep-ph/9512328]
}
\preref\Byckling{E. Byckling and K. Kajantie, {\it Particle Kinematics\/}
(Wiley, 1973)}
\preref\OneloopSplitB{
Z.\ Bern, V.\ Del Duca and C.\ R.\ Schmidt,
Phys.\ Lett.\ B445:168 (1998) [hep-ph/9810409]\semi
Z.\ Bern, V.\ Del Duca, W.\ B.\ Kilgore and C.\ R.\ Schmidt,
Phys.\ Rev.\ D60:116001 (1999) [hep-ph/9903516]
}
\preref\OneloopSplitA{
D.\ A.\ Kosower and P.\ Uwer,
Nucl.\ Phys.\ B563:477 (1999) [hep-ph/9903515]
}

\preref\SWI{M. T. Grisaru, H. N. Pendleton and P. van
Nieuwenhuizen,
Phys. Rev. D15:996 (1977)\semi
M. T. Grisaru and H.N. Pendleton, Nucl. Phys. B124:81 (1977)}
\preref\UseSWI{
S. J. Parke and T. Taylor, Phys. Lett. B157:81 (1985)\semi
Z. Kunszt, Nucl. Phys. B271:333 (1986)}
\preref\Lewin{L. Lewin, {\it Polylogarithms and associated functions\/}
(North-Holland,1981)}

\preref\ParkeTaylor{S. J. Parke and T. R. Taylor,
Phys.\ Rev.\ Lett.\ 56:2459 (1986)}

\preref\Nair{%
V.\ P.\ Nair,
Phys.\ Lett.\ B214:215 (1988)%
}

\preref\WittenTopologicalString{
E.\ Witten,
hep-th/0312171%
}

\preref\CSW{
F.\ Cachazo, P.\ Svrcek and E.\ Witten,
hep-th/0403047%
}

\preref\RSV{
R.\ Roiban, M.\ Spradlin and A.\ Volovich,
JHEP {0404}:012 (2004) [hep-th/0402016]\semi
R.\ Roiban and A.\ Volovich,
hep-th/0402121\semi
R.\ Roiban, M.\ Spradlin and A.\ Volovich,
hep-th/0403190%
}

\preref\Gukov{
S.\ Gukov, L.\ Motl and A.\ Neitzke,
hep-th/0404085%
}

\preref\Berkovits{
N.\ Berkovits,
hep-th/0402045\semi
N.\ Berkovits and L.\ Motl,
JHEP {0404}:056 (2004)
[hep-th/0403187]%
}

\preref\Siegel{
W.\ Siegel,
hep-th/0404255%
}

\preref\Khoze{
G.\ Georgiou and V.\ V.\ Khoze,
hep-th/0404072%
}

\preref\OtherGoogly{
C.\ J.\ Zhu,
JHEP {0404}:032 (2004)
[hep-th/0403115]\semi%
J.\ B.\ Wu and C.\ J.\ Zhu,
hep-th/0406085.
}

\preref\Popov{
A.\ D.\ Popov and C.\ Saemann,
hep-th/0405123%
}

\loadfourteenpoint
\noindent\nopagenumbers
[hep-th/0406175] \hfill{Saclay/SPhT--T04/067\ \ \hfil NSF-KITP-04-58}

\leftlabelstrue
\vskip -0.7 in
\Title{Next-to-Maximal Helicity Violating Amplitudes in Gauge Theory}
\vskip 10pt

\baselineskip17truept
\centerline{David A. Kosower}
\baselineskip12truept
\centerline{\it Service de Physique Th\'eorique${}^{\natural}$}
\centerline{\it CEA--Saclay}
\centerline{\it F-91191 Gif-sur-Yvette cedex, France}
\centerline{\tt kosower@spht.saclay.cea.fr}

\vskip 0.2in\baselineskip13truept

\vskip 0.5truein
\centerline{\bf Abstract}
{\narrower 

Using the novel diagrammatic rules recently proposed by Cachazo, Svrcek, and
Witten, I give a compact, manifestly Lorentz-invariant form for
tree-level gauge-theory amplitudes with three opposite helicities.
}
\vskip 0.3truein


\vfill
\vskip 0.1in
\noindent\hrule width 3.6in\hfil\break
\noindent
${}^{\natural}$Laboratory of the
{\it Direction des Sciences de la Mati\`ere\/}
of the {\it Commissariat \`a l'Energie Atomique\/} of France.\hfil\break

\Date{}

\line{}

\baselineskip17pt
%

\section{Introduction}
\vskip 10pt

\def\SU{S\hskip -0.7pt{}U}
The computation of amplitudes in gauge theories is important to
future physics analyses at colliders.  Tree-level amplitudes
in perturbative QCD, for example, provide the leading-order
approximation to multi-jet
processes at hadron colliders.  At tree level, all-gluon
amplitudes in pure $\SU(N)$ gauge theory
 are in fact identical to those in the ${\cal N}=4$
supersymmetric theory, since only gluons can appear as interior lines.

This hidden supersymmetry manifests itself in the vanishing of amplitudes
for certain helicity configurations, namely those in which all or all but one
gluon helicities are identical~[\use\SWI]. (I follow the usual convention where all 
momenta are taken to be outgoing; recall that flipping a momentum from
outgoing to incoming also flips the helicity.)  These vanishings are
expressed in two of the three Parke--Taylor equations~[\use\ParkeTaylor],
$$\eqalign{
{\cal A}_n^{\tree}(k_1^+,k_2^+,\ldots,k_n^+) &= 0,\cr
{\cal A}_n^{\tree}(k_1^-,k_2^+,\ldots,k_n^+) &= 0.\cr
}\anoneqn$$
The next amplitude in this sequence, with two opposite-helicity gluons,
is the maximally helicity-violating amplitude that does not vanish, and
is conventionally called the MHV amplitude.  It has a simple form,
recalled below, given
by the third Parke--Taylor equation.  Continuing in the sequence,
we find amplitudes with three opposite helicity gluons, which I will
call `next-to-MHV' or NMHV for short.  These are the subject of this paper.

The ${\cal N}=4$ supersymmetric theory is of great interest for
a number of reasons, especially its links with string theories.
Witten has recently proposed [\use\WittenTopologicalString]
a novel link between a twistor-space topological string
theory and the amplitudes of the ${\cal N}=4$
supersymmetric gauge theory.  This proposal generalizes
Nair's earlier construction~[\use\Nair] of MHV amplitudes.
A number of authors have investigated
issues connected with the derivation of gauge-theory amplitudes from the
topological string theory~[\use\RSV,\use\Gukov] as well as alternative
approaches~[\use\Berkovits,\use\Siegel] and related 
issues~[\use\Khoze,\use\OtherGoogly].
Based on investigations in this
string theory, Cachazo, Svrcek, and Witten (CSW) proposed [\use\CSW]
a novel construction
of tree-level gauge-theory amplitudes.  It expresses
any amplitude in terms of propagators and basic `vertices' which are
off-shell continuations of the Parke-Taylor amplitudes.  The construction
makes manifest the factorization on multi-particle poles.

Cachazo et.~al.~used their construction to give a simple form
for an amplitude in the NMHV class.  (See eqn.~(3.7) of ref.~[\use\CSW].)
Their form extends straightforwardly
to all NMHV amplitudes, and is sufficient for numerical computations.
However, although it is Lorentz invariant, the invariance is not manifest
because of apparent (spinor) poles involving an external reference momentum.
For the same reason, the CSW form is not convenient for feeding into 
the unitarity machinery [\use\UnitarityMachinery]
in order to compute loop 
amplitudes~[\use\UnitarityCalculations]\footnote{${}^\dagger$}{As noted 
in ref.~[\use\CSW], their form can be made Lorentz-invariant by choosing 
the reference momentum to be one of the momenta of the process after some
manipulation; but this still leaves undesirable denominators,
ones that do not reconstruct
propagators in the unitarity-based sewing method.}.  The purpose
of this paper is to transform the CSW form into one which is manifestly
Lorentz-invariant, and suitable for use as a building block in computing
loop amplitudes.  In the next section, I review the off-shell
continuation needed to formulate the CSW construction, which I discuss
in section~\use\CSWsection.  I compute the NMHV amplitude in 
section~\use\NMHVsection.

\vskip 10pt
\section{Off-Shell Continuations}
\vskip 10pt

It is convenient to write the full tree-level amplitude using
a color decomposition~[\use\Color],
$$
{\cal A}_n^\tree(\{k_i,\lambda_i,a_i\}) = 
\sum_{\sigma \in S_n/Z_n} \Tr(T^{a_{\sigma(1)}}\cdots T^{a_{\sigma(n)}})\,
A_n^\tree(\sigma(1^{\lambda_1},\ldots,n^{\lambda_n}))\,,
\eqn\TreeColorDecomposition$$
where $S_n/Z_n$ is the group of non-cyclic permutations
on $n$ symbols, and $j^{\lambda_j}$ denotes the $j$-th momentum
and helicity $\lambda_j$.  The notation $j_1+j_2$ appearing below will denote
the sum of momenta, $k_{j_1}+k_{j_2}$.
I use the normalization $\Tr(T^a T^b) = \delta^{ab}$.  The
color-ordered amplitude $A_n$ is invariant under
a cyclic permutation of its arguments.  It is the object we will calculate
directly.

\def\vo{\vphantom{1}}
Using spinor products~[\use\Spinor], 
the third Parke--Taylor equation takes the simple form,
$$
A_n^\tree(1^+,\ldots,m_1^{-},(m_1\!+\!1)\vo^{+},\ldots,m_2^{-},
          (m_2\!+\!1)\vo^{+},\ldots,n^+) = 
  i {\spa{m_1}.{m_2}^4\over \spa1.2\spa2.3\cdots \spa{(n\!-\!1)}.n\spa{n}.1}.
\anoneqn$$

The CSW construction [\use\CSW] builds amplitudes out of building
blocks which are off-shell continuations of the Parke--Taylor amplitudes.
We can obtain an off-shell formulation which is equivalent to the
CSW one (but slightly more convenient for explicit calculations) by
considering first the off-shell continuation of a gluon polarization vector.
In sewing an off-shell gluon carrying momentum $K$, we will want to sum over
all (physical) polarization states.  We can do this via the identity,
$$
\sum_\sigma \pol^{(\sigma)}_\mu(K,q) \pol^{*(\sigma)}_\nu(K,q) = 
-g_{\mu\nu} + {K_\mu q_\nu + q_\mu K_\nu\over q\cdot K},
\eqn\PolarizationSum$$
where $q$ is the reference or light-cone vector, satisfying $q^2=0$.

\def\proj{\flat}
Observe that we can always decompose the off-shell momentum
$K$ into a sum of two massless momenta, where one is proportional to $q$,
$$
K = k^\proj + \eta(K) q.
\anoneqn$$
The constraint $(k^\proj)^2 = 0$ yields
$$
\eta(K) =  {K^2 \over 2 q\cdot K}.
\anoneqn$$
Of course, if $K$ goes on shell, $\eta$ vanishes.  Also, if two
off-shell vectors sum to zero, $K_1+K_2=0$, then so do the corresponding
$k^\proj$s.

Noting that $q\cdot K = q\cdot k^\proj$, 
we can then rewrite eqn.~(\use\PolarizationSum) as follows,
$$\eqalign{
\sum_\sigma \pol^{(\sigma)}_\mu(K,q) \pol^{*(\sigma)}_\nu(K,q) &= 
-g_{\mu\nu} + {k_{*\mu} q_\nu + q_\mu k_{*\nu}\over q\cdot k^\proj}
+{2\eta(K)q_\mu q_\nu\over q\cdot k^\proj}
\cr &= \sum_\sigma \pol^{(\sigma)}_\mu(k^\proj,q) \pol^{*(\sigma)}_\nu(k^\proj,q)
+{2\eta(K)q_\mu q_\nu\over q\cdot k^\proj}
\cr &= \sum_\sigma \pol^{(\sigma)}_\mu(k^\proj,q) \pol^{*(\sigma)}_\nu(k^\proj,q)
+{K^2\over (q\cdot K)^2} q_\mu q_\nu.
}\eqn\Decomposition$$
In this expression, $\pol(k^\proj,q)$ is of course just the polarization
vector for a massless momentum, and so can be expressed in
terms of spinor products.  The power of $K^2$ in the second term will
cancel the $1/K^2$ in the propagator, leading to an additional
contribution to the four-point vertex.  One can formulate a light-cone
version of the recurrence relations using such a modified four-point
vertex, and retaining only the first term for the 
gluon propagator.  This leads to the simple rule of continuing
an amplitude off-shell by replacing
$\pol_\mu(k,q)\rightarrow \pol_\mu(k^\proj,q)$.  For MHV amplitudes, this
amounts to the prescription,
$$
\spa{j}.{j'}\rightarrow \spa{\smash{j^\proj}}.{j'},
\anoneqn$$
when $k_j$ is taken off shell.

The choice of the momentum $q$ is equivalent to the choice of the
constant spinor $\eta$ in ref.~[\use\CSW].  The continuation given
there amounts to taking
$$
\spa{j}.{j'} \rightarrow \spb{q}.{j}\spa{j}.{j'} \rightarrow
\sandpp{q}.{\Ksl_j}.{j'}; 
\anoneqn$$
but this is just equal to
$$
\sandpp{q}.{\ksl_j^\proj}.{j'} = \spb{q}.{\smash{j^\proj}}\spa{\smash{j^\proj}}.{j'}.
\anoneqn$$
(The extra factors of $\spb{q}.{\smash{j^\proj}}$ 
present in the CSW construction
cancel when sewing vertices into an on-shell amplitude.)

In the notation employed here, an amplitude is manifestly Lorentz invariant
(or equivalently manifestly gauge invariant) when it is manifestly free
of $q$, whether present explicitly or implicitly via $k^\proj$s.

\section{Amplitudes from MHV Building Blocks}
\tagsection\CSWsection
\vskip 10pt

The CSW construction replaces ordinary Feynman diagrams with diagrams
built out of MHV vertices and ordinary propagators.  Each vertex has
exactly two lines carrying negative helicity (which may be on or off
shell), and any number of lines carrying positive helicity.  The propagator
takes the simple form $i/K^2$, because the physical state 
projector~(\use\PolarizationSum) is now effectively supplied by the
vertices.  The simplest vertex is an amplitude with one leg taken
off shell,
\def\ks#1{\{#1\}}
$$
A_n^\tree(1^+,\ldots,m_1^{-},(m_1\!+\!1)\vo^{+},\ldots,m_2^{-},
          (m_2\!+\!1)\vo^{+},\ldots,(n\!-\!1)\vo^+,(-K_{1\cdots (n-1)})\vo^+),
\anoneqn$$
where $K_{j\cdots l} \equiv k_j+\cdots+k_l$.

It will be convenient to denote the projected
$k^\proj$ momentum built out of $-K_{1\cdots n}$ by $\ks{1\cdots n}$, for example 
$\spa{j}.{k^\proj(-K_{1\cdots n},q)} = \spa{j}.{\ks{1\cdots n}}$.  Because
$$\eqalign{
k^\proj(K+{k'}\vphantom{k}^\proj,q) &= K+K' - \eta(K') q - \eta(K+k'_*) q
\cr &= K+K' - \biggl[ \eta(K')
                      + {K^2 + 2 K\cdot K'
                      - \eta(K') 2q\cdot K\over 2q\cdot(K+K')}\biggr] q
\cr &= K+K' - {K^2 + 2 K\cdot K'
                      + \eta(K') 2q\cdot K'\over 2q\cdot(K+K')} q
\cr &= k^\proj(K+K',q),
}\eqn\OffShellInOffShell$$
it does not matter whether we feed in the original off-shell
momentum or the corresponding massless projection.

The simplest vertices then have the explicit expression,
$$\eqalign{
A_n^\tree&(1^+,\ldots,m_1^{-},(m_1\!+\!1)\vo^{+},\ldots,m_2^{-},
          (m_2\!+\!1)\vo^{+},\ldots,(-K_{1\cdots (n-1)})\vo^+) =\cr
&\hskip 10mm  {i\spa{m_1}.{m_2}^4\over \spa1.2\spa2.3\cdots 
      \spa{(n\!-\!1)}.{\ks{1\cdots(n\!-\!1)}}\spa{\ks{1\cdots(n\!-\!1)}}.1},\cr
A_n^\tree&(1^+,\ldots,m_1^{-},(m_1\!+\!1)\vo^{+},\ldots,
          (-K_{1\cdots (n-1)})\vo^-) =\cr
&\hskip 10mm {i\spa{m_1}.{\ks{1\cdots(n\!-\!1)}}^4\over \spa1.2\spa2.3\cdots 
      \spa{(n\!-\!1)}.{\ks{1\cdots(n\!-\!1)}}\spa{\ks{1\cdots(n\!-\!1)}}.1},
}\eqn\Vertices$$

\topinsert\LoadFigure\CSWTermFigure
{\baselineskip 13 pt
\noindent\narrower
A term in the CSW representation of an NMHV amplitude.  The black dot represents
the multiparticle pole multiplying the two on-shell amplitudes.
}  {\epsfxsize 3.5 truein}{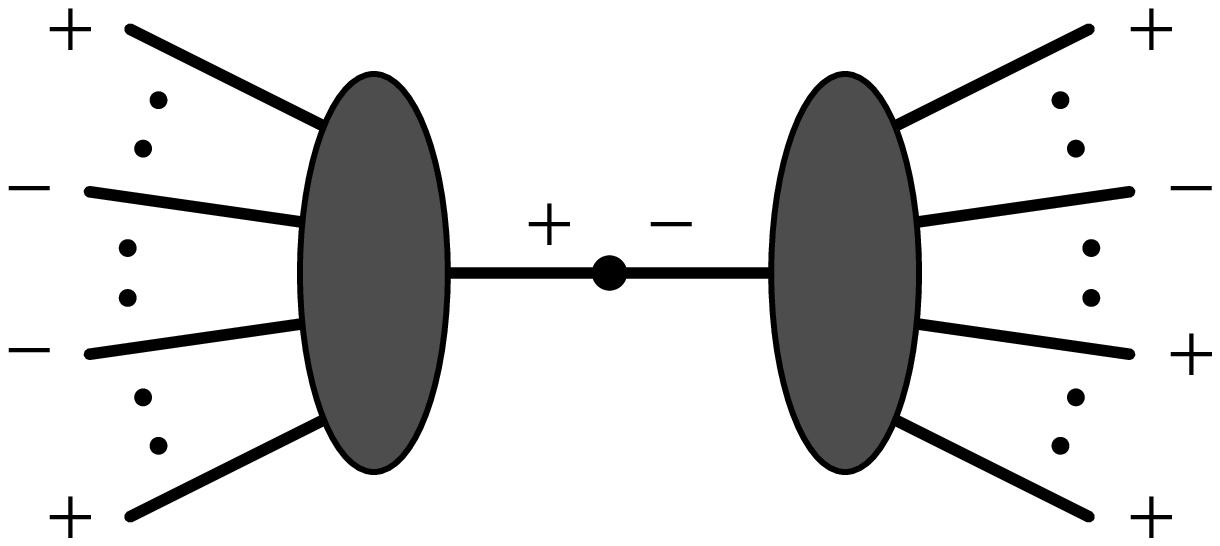}{}\endinsert

The CSW rules then instruct us to write down all tree diagrams with
MHV vertices, subject to the constraints that each vertex have exactly
two negative-helicity gluons attached, and that each propagator connect
legs of opposite helicity.
 For amplitudes with two negative-helicity
gluons, the vertex with all legs taken on shell is then the amplitude; for
amplitudes with three negative-helicity gluons, we must write down all
diagrams with two vertices.  One of the vertices has two of the external
negative-helicity gluons attached to it, while the other has only one.
An example of such a diagram is shown in 
fig.~\use\CSWTermFigure.

This leads to the following
form (without loss of generality, we may take the first leg to have
negative helicity),
\def\cm#1{\kern -4pt#1\kern 1pt}
\def\pI{{\phantom1}}
$$\eqalign{
A_n^\tree&(1^-,2^+,\ldots,\cm,
     m_2^-,(m_2+1)^+\cm,\ldots,m_3^-,
     (m_3+1)_\pI^+\cm,\ldots,n^+) = \vphantom{\sum}\cr
&\smash{ \mathop{\sum_{j_1=m_2+1}^{m_3} \sum_{j_2=2}^{m_2}}_{%
              (j_1,j_2)\neq (m_2+1,m_2)} }
           {i\over s_{j_2\cdots (j_1\!-\!1)}}
   A_{j_1-j_2+1}^\tree(j_2^+,\ldots,m_2^-,\ldots,(j_1-1)^+,
                       \ks{j_2\cdots(j_1-1)}^-)
\cr &\hphantom{%
\mathop{\sum_{j_1=m_2+1}^{n+1} \sum_{j_2=2}^{m_2}}_{%
              (j_1,j_2)\neq (m_2+1,m_2)}
           {i\over s_{j_2\cdots (j_1\!-\!1)}}  A_{n-j_1+j_2+1}()  }\hskip 3mm
\cr &\hphantom{\sum_{j=2}^{m_2-1} {1\over s_{1\cdots j}}}
       \vphantom{\sum} \hskip 3mm\times
   A_{n-j_1+j_2+1}^\tree(1^-,2^+,\ldots,(j_2-1)^+,
                         \ks{1\cdots (j_2\!-\!1);j_1\cdots n}^+,
           j_1^+,\ldots,m_3^-,\ldots,n^+)
\cr &+\hskip -2mm
\mathop{\sum_{j_1=m_3+1}^{n+1} \sum_{j_2=m_2+1}^{m_3}}_{%
              (j_1,j_2)\neq (m_3+1,m_3)}
    {i\over s_{j_2\cdots (j_1\!-\!1)}}
 A_{j_1-j_2+1}^\tree(j_2^+,\ldots,m_3^-,\ldots,(j_1-1)^+,
                     \ks{j_2\cdots (j_1\!-\!1)}^-)
\cr &\hphantom{+\sum_{j=m_2}^{m_3-1} {1\over s_{1\cdots j}}}
       \hskip 3mm\times
 A_{n-j_1+j_2+1}^\tree(1^-,2^+,\ldots,
    m_2^-,\ldots,(j_2-1)^+,
           \ks{1\cdots (j_2\!-\!1);j_1\cdots n}_{\phantom1}^+\kern -4pt,
           j_1^+,\ldots,n^+)
\cr &+\smash{ \mathop{\sum_{j_1=2}^{m_2}\sum_{j_2=m_3+1}^{n+1}}_{%
              (j_1,j_2)\neq(2,n+1)} }
   {i\over s_{j_1\cdots (j_2-1)}}
   A_{j_2-j_1+1}^\tree(j_1^+,\ldots,m_2^-,\ldots,
         m_3^-,\ldots,(j_2-1)^+,\ks{j_1\cdots (j_2\!-\!1)}^+)
\cr &\hphantom{+\sum_{j_1=2}^{m_2}\sum_{j_2=m_3+1}^{n+1} 
{1\over s_{j_1\cdots (j_2-1)}}}
      \hskip 3mm\hphantom{A_{j_2-j_1+1}()}
\cr &\hphantom{+\sum_{j=m_2}^{m_3-1} {1\over s_{1\cdots j}}}\hskip 3mm\times
   A_{n-j_2+j_1+1}^\tree(j_2^+,\ldots,n^+,
                   1^-,2^+,\ldots,(j_1-1)^+,\ks{1\cdots (j_1\!-\!1);j_2\cdots n}^-)
}\eqn\ThreeMinusGeneral$$
where $s_{j\cdots l} = K_{j\cdots l}^2 \equiv (k_j+\cdots+k_l)^2$, 
and all indices are to be understood
$\mod n$.  The three double sums correspond to the three different choices
($(1,m_2)$, $(m_2,m_3)$, or $(m_3,1)$) we
can make for the pair of negative-helicity gluons which enter the same MHV vertex.
In the next section, I will evaluate this expression explicitly.

\section{Next-to-MHV Amplitudes}
\tagsection\NMHVsection
\vskip 10pt

Begin the evaluation of eqn.~(\use\ThreeMinusGeneral) by substituting
the explicit forms of the vertices~(\use\Vertices), and then remove
an overall factor of 
$i(\spa1.2\spa2.3\cdots \spa{n}.1)^{-1}$.

A generic term in the second double sum then has the form,
$$\eqalign{
&{\spa1.{m_2}^4 \spa{(j_1-1)}.{j_1} \spa{(j_2-1)}.{j_2}
 \spa{m_3}.{\ks{j_1\cdots(j_2-1)}}^4
 \over \spa{(j_2-1)}.{\ks{j_1\cdots (j_2-1)}}
       \spa{\ks{j_1\cdots (j_2-1)}}.{j_1}}
\cr&\times {1\over
       \spa{(j_1-1)}.{\ks{j_2\cdots (j_1-1)}}
       \spa{\ks{j_2\cdots (j_1-1)}}.{j_2} s_{j_2\cdots (j_1-1)}}.
}\eqn\GenericI$$
where $\ks{j_1\cdots (j_2\!-\!1)} \equiv
\ks{1\cdots (j_2\!-\!1);j_1\cdots n}$.
We can use momentum conservation, followed by
the Schouten identity 
$\spa{a}.{b} \spa{c}.{d} = \spa{a}.{d}\spa{c}.{b}+\spa{a}.{c}\spa{b}.{d}$
to rewrite this as,
$$\eqalign{
&-{\spa1.{m_2}^4  \spa{m_3}.{\ks{j_1\cdots(j_2-1)}}^2
  \over s_{j_1\cdots (j_2-1)}}
\biggl({\spa{m_3}.{j_1}\over\spa{\ks{j_1\cdots(j_2-1)}}.{j_1}}
-{\spa{m_3}.{(j_1-1)}\over\spa{\ks{j_1\cdots(j_2-1)}}.{(j_1-1)}}
  \biggr)
\cr &\hskip 5mm\times
\biggl({\spa{m_3}.{j_2}\over\spa{\ks{j_1\cdots(j_2-1)}}.{j_2}}
-{\spa{m_3}.{(j_2-1)}\over\spa{\ks{j_1\cdots(j_2-1)}}.{(j_2-1)}}
  \biggr)
}\eqn\GenericII$$

If we now gather all terms in a double sum containing $\spa{m_3}.{j_1}$
and $\spa{m_3}.{j_2}$ (for generic values of $j_1$ and $j_2$), we find
$$\eqalign{
&-\spa1.{m_2}^4  \spa{m_3}.{j_1} \spa{m_3}.{j_2}
\cr &\times
\biggl(
{\spa{m_3}.{\ks{j_1\cdots(j_2-1)}}^2\over\spa{\ks{j_1\cdots(j_2-1)}}.{j_1}
\spa{\ks{j_1\cdots(j_2-1)}}.{j_2}
s_{j_1\cdots (j_2-1)}}
\cr &\hphantom{\times\biggl()}
-{\spa{m_3}.{\ks{(j_1+1)\cdots(j_2-1)}}^2
  \over\spa{\ks{(j_1+1)\cdots(j_2-1)}}.{j_1}
       \spa{\ks{(j_1+1)\cdots(j_2-1)}}.{j_2}
       s_{(j_1+1)\cdots (j_2-1)}}
\cr &\hphantom{\times\biggl()}
-{\spa{m_3}.{\ks{j_1\cdots j_2}}^2\over\spa{\ks{j_1\cdots j_2}}.{j_1}
  \spa{\ks{j_1\cdots j_2}}.{j_2}
  s_{j_1\cdots j_2}}
\cr &\hphantom{\times\biggl()}
+{\spa{m_3}.{\ks{(j_1+1)\cdots j_2}}^2
  \over\spa{\ks{(j_1+1)\cdots j_2}}.{j_1}
   \spa{\ks{(j_1+1)\cdots j_2}}.{j_2}
   s_{(j_1+1)\cdots j_2}}
 \biggr)
}\eqn\StepI$$

Multiply and divide by $\spa{j_1}.{j_2}$, and use the Schouten
identity again to split denominators,
$$\eqalign{
&-{\spa1.{m_2}^4 \spa{m_3}.{j_1} \spa{m_3}.{j_2}\over \spa{j_1}.{j_2}} 
\cr &\hskip 5mm\times \biggl[
  \spa{m_3}.{j_1} \biggl(
   -{\spa{m_3}.{\ks{j_1\cdots (j_2-1)}}\over
      \spa{j_1}.{\ks{j_1\cdots (j_2-1)}} s_{j_1\cdots (j_2-1)}}
   +{\spa{m_3}.{\ks{(j_1+1)\cdots (j_2-1)}}\over
      \spa{j_1}.{\ks{(j_1+1)\cdots (j_2-1)}} s_{(j_1+1)\cdots (j_2-1)}}
      \biggr)
\cr &\hskip 5mm\hphantom{\times\biggl[]}
  +\spa{m_3}.{j_1} \biggl(
   +{\spa{m_3}.{\ks{j_1\cdots j_2}}\over
      \spa{j_1}.{\ks{j_1\cdots j_2}} s_{j_1\cdots j_2}}
   -{\spa{m_3}.{\ks{(j_1+1)\cdots j_2}}\over
      \spa{j_1}.{\ks{(j_1+1)\cdots j_2}} s_{(j_1+1)\cdots j_2}}
      \biggr)
\cr &\hskip 5mm\hphantom{\times\biggl[]}
  +\spa{m_3}.{j_2} \biggl(
   {\spa{m_3}.{\ks{j_1\cdots (j_2-1)}}\over
      \spa{j_2}.{\ks{j_1\cdots (j_2-1)}} s_{j_1\cdots (j_2-1)}}
   -{\spa{m_3}.{\ks{j_1\cdots j_2}}\over
      \spa{j_2}.{\ks{j_1\cdots j_2}} s_{j_1\cdots j_2}}
    \biggr)
\cr &\hskip 5mm\hphantom{\times\biggl[]}
  +\spa{m_3}.{j_2} \biggl(
   -{\spa{m_3}.{\ks{(j_1+1)\cdots (j_2-1)}}\over
      \spa{j_2}.{\ks{(j_1+1)\cdots (j_2-1)}} s_{(j_1+1)\cdots (j_2-1)}}
   +{\spa{m_3}.{\ks{(j_1+1)\cdots j_2}}\over
      \spa{j_2}.{\ks{(j_1+1)\cdots j_2}} s_{(j_1+1)\cdots j_2}}
    \biggr)
           \biggr]
}\eqn\StepII$$

We must next use a partial-fractioning identity,
$$\eqalign{
{\spa{m}.{\ks{K,k_j}}\over \spa{j}.{\ks{K,k_j}} (K+k_j)^2}
&= {\sand{m}.{\ks{K,k_j}}.j\over 2\,k_j\cdot{\ks{K,k_j}} (K+k_j)^2}
\cr &= {\sand{m}.{\ks{K,k_j}}.j 2\,q\cdot (K+k_j)\over 
      [-4\,q\cdot (K+k_j) k_j\cdot K+2\,(K+k_j)^2 q\cdot k_j] (K+k_j)^2}
\cr &= -{\sand{m}.{\Ksl}.j\over K^2 (K+k_j)^2}
+{\sand{m}.{\ks{K}}.j\over 2\,k_j\cdot{\ks{K}} K^2 }
\cr &= -{\sand{m}.{\Ksl}.j\over K^2(K+k_j)^2}
+{\spa{m}.{\ks{K}} \over \spa{j}.{\ks{K}}K^2 }
}\eqn\PFIdentity$$
which in turn relies on the identity
$$
q\cdot (K+k_j)\,k_j\cdot\ks{K,k_j} = q\cdot K\,k_j\cdot\ks{K}.
\anoneqn$$

Using the identity~(\use\PFIdentity),
 we can rewrite eqn.~(\use\StepII); doing so,
separating out terms without a $j_1\parallel j_2$ singularity, and
collecting terms, we obtain
$$\eqalign{
& -{\spa1.{m_2}^4 \spa{m_3}.{j_1} \spa{m_3}.{j_2}\over\spa{j_1}.{j_2}}
\cr &\hskip 5mm\times\biggl[
  {\spa{m_3}.{j_1} \spa{m_3}.{j_2} \spb{j_1}.{j_2}\over
      s_{j_1\cdots j_2} s_{(j_1+1)\cdots j_2}}
  +{\spa{m_3}.{j_2} \spa{m_3}.{j_1} \spb{j_1}.{j_2}\over
      s_{j_1\cdots (j_2-1)} s_{j_1\cdots j_2}}
\cr &\hskip 5mm\hphantom{\times\biggl[]}
  +{\spa{m_3}.{j_1} \sand{m_3}.{\Ksl_{(j_1+1)\cdots (j_2-1)}}.{j_1}
    \over s_{j_1\cdots (j_2-1)}  s_{(j_1+1)\cdots (j_2-1)}
        s_{j_1\cdots j_2}  s_{(j_1+1)\cdots j_2}}\Bigl( 
     s_{j_1\cdots j_2} s_{(j_1+1)\cdots j_2}
     - s_{j_1\cdots (j_2-1)}  s_{(j_1+1)\cdots (j_2-1)}
     \Bigr)
\cr &\hskip 5mm\hphantom{\times\biggl[]}
  +{\spa{m_3}.{j_2} \sand{m_3}.{\Ksl_{(j_1+1)\cdots (j_2-1)}}.{j_2}
    \over s_{(j_1+1)\cdots (j_2-1)} s_{(j_1+1)\cdots j_2}
         s_{j_1\cdots (j_2-1)} s_{j_1\cdots j_2}} \Bigl(
     s_{(j_1+1)\cdots (j_2-1)} s_{(j_1+1)\cdots j_2}
     - s_{j_1\cdots (j_2-1)} s_{j_1\cdots j_2}
     \Bigr)
           \biggr].
}\eqn\StepIII$$
Note that all $q$ dependence has disappeared.

Rewriting the differences of invariants,
using the Schouten identity twice (on the sandwich product
$\sand{m_3}.{\Ksl_{(j_1+1)\cdots (j_2-1)}}.{j_1}\*
 \sand{j_2}.{\Ksl_{(j_1+1)\cdots (j_2-1)}}.{j_2}$, and then
on the product $\spa{m_3}.{j_1}\* \sand{j_2}.{\Ksl_{(j_1+1)\cdots (j_2-1)}}.{j_1}$),
and combining terms, we obtain two final equivalent forms,
$$\eqalign{
& \spa{1}.{m_2}^4 
\biggl(
  {\sandmp{m_3}.{\Ksl_{(j_1+1)\cdots (j_2-1)}\ksl_{j_1}}.{m_3} 
   \sandmp{m_3}.{\Ksl_{j_1\cdots j_2}\ksl_{j_2}}.{m_3}\over
     s_{j_1\cdots (j_2-1)}  s_{(j_1+1)\cdots (j_2-1)}
        s_{j_1\cdots j_2} } 
\cr & \hphantom{ \spa{1}.{m_2}^4 \biggl() }
  +{\sandmp{m_3}.{\Ksl_{j_1\cdots j_2}\ksl_{j_1}}.{m_3}
   \sandmp{m_3}.{\Ksl_{(j_1+1)\cdots (j_2-1)}\ksl_{j_2}}.{m_3}\over
     s_{(j_1+1)\cdots (j_2-1)} s_{j_1\cdots j_2}  s_{(j_1+1)\cdots j_2}}
\biggr)
\cr &= \spa{1}.{m_2}^4 
\biggl(
  {\sandmp{m_3}.{\Ksl_{j_1\cdots j_2}\ksl_{j_1}}.{m_3} 
   \sandmp{m_3}.{\Ksl_{j_1\cdots j_2}\ksl_{j_2}}.{m_3}\over
     s_{j_1\cdots (j_2-1)}  s_{(j_1+1)\cdots j_2}
        s_{j_1\cdots j_2} }
\cr & \hphantom{ \spa{1}.{m_2}^4 \biggl() }
  +{\sandmp{m_3}.{\Ksl_{(j_1+1)\cdots (j_2-1)}\ksl_{j_1}}.{m_3}
   \sandmp{m_3}.{\Ksl_{(j_1+1)\cdots (j_2-1)}\ksl_{j_2}}.{m_3}\over
     s_{(j_1+1)\cdots (j_2-1)} s_{j_1\cdots (j_2-1)}  s_{(j_1+1)\cdots j_2}}
\biggr).
}\anoneqn$$

The `boundary' terms in the double sums will not contribute all
of the terms in eqn.~(\use\StepI); but the remaining $q$ dependence
cancels against that in the other double sums.  The resulting
computation leads one to define,
$$\eqalign{
H&(m_0,m_1,m_2) =\cr
& -\spa{m_1}.{m_2}^4 
\sum_{j_1=m_2+1}^{m_0}\mathop{{\sum}'}_{j_2=m_0+1}^{m_1}
\biggl(
  {\sandmp{m_0}.{\Ksl_{j_1\cdots j_2}\ksl_{j_1}}.{m_0} 
   \sandmp{m_0}.{\Ksl_{j_1\cdots j_2}\ksl_{j_2}}.{m_0}\over
     s_{j_1\cdots (j_2-1)}  s_{(j_1+1)\cdots j_2}
        s_{j_1\cdots j_2} }
\cr & \hphantom{\spa{m_1}.{m_2}^4 %
                \mathop{\sum_{j_1=m_2+1}^{m_0-1}%
                        \sum_{j_2=m_0+1}^{m_1-1}}_{(j_1,j_2)\neq (m-1,m+1)}%
                \biggl()}
  +{\sandmp{m_0}.{\Ksl_{(j_1+1)\cdots (j_2-1)}\ksl_{j_1}}.{m_0}
   \sandmp{m_0}.{\Ksl_{(j_1+1)\cdots (j_2-1)}\ksl_{j_2}}.{m_0}\over
     s_{(j_1+1)\cdots (j_2-1)} s_{j_1\cdots (j_2-1)}  s_{(j_1+1)\cdots j_2}}
\biggr)
\cr &\hskip 2mm -\spa{m_0}.{m_1}\spa{m_1}.{m_2}\spa{m_2}.{m_0}
   \mathop{{\sum}'}_{j=m_0+1}^{m_1} {1\over s_{(j+1)\cdots m_2} s_{j\cdots m_2}}
\cr &\hskip 2mm\hphantom{ \sum_{j=m_0+1}^{m_1} }\hskip 5mm\times
 \biggl( \spa{m_2}.{m_0}^2 \sandmp{m_1}.{\Ksl_{(j+1)\cdots m_2}\ksl_j}.{m_1}
        +\spa{m_1}.{m_2}^2  \sandmp{m_0}.{\Ksl_{(j+1)\cdots m_2}\ksl_j}.{m_0}
\cr &\hskip 2mm\hphantom{ \sum_{j=m_0+1}^{m_1} \times\biggl()}\hskip 5mm
        -\spa{m_1}.{m_2}\spa{m_2}.{m_0} 
              \sandmp{m_0}.{\Ksl_{(j+1)\cdots m_2}\ksl_j}.{m_1}
 \biggr)
\cr &\hskip 2mm +\delta_{m_0\neq m_2+1}
              {\spa{m_0}.{m_1}^2 \spa{m_1}.{m_2}^2 \spa{m_2}.{m_0}^2
                  \over s_{(m_2+1)\cdots m_0}}
\cr &\hskip 2mm -\delta_{m_1+1,m_2} 
            \spa{m_0}.{m_1} \spa{m_1}.{m_2}^2 \spa{m_2}.{m_0}
               \sand{m_0}.{\Ksl_{m_1\cdots m_2}}.{(m_2\!+\!1)\vphantom{m_2}}
\cr &\hskip 2mm\hphantom{ -\delta_{m_1+1,m_2} \spa{m_0}.{m_1} }\times
             {\bigl(\spa{m_1}.{m_2} \spa{m_0}.{(m_2\!+\!1)}
              -\spa{m_0}.{m_1} \spa{m_2}.{(m_2\!+\!1)}\bigr)
             \over s_{m_1\cdots m_2} s_{m_2 (m_2+1)}},
}\eqn\Hdefinition$$
where the prime signifies that any term with a vanishing denominator
is to be omitted (as does the $\delta_{m_0\neq m_2+1}$), and
where the sums over indices are understood to run along the cyclic order; that is,
for example,
$$
\sum_{j=n-4}^3 \equiv \sum_{j={(n-4)\cdots n,1\cdots3}}.
\anoneqn$$

With this function, we can then write the desired amplitude in the compact form,
$$\eqalign{
A_n^\tree&(1^-,2^+,\ldots,\cm,
     m_2^-,(m_2+1)^+\cm,\ldots,m_3^-,
     (m_3+1)_\pI^+\cm,\ldots,n^+) = \cr
&\hskip 15mm
{i(H(1,m_2,m_3)+H(m_2,m_3,1)+H(m_3,1,m_2))
 \over \spa1.2\spa2.3\cdots \spa{(n\!-\!1)}.n \spa{n}.1}.
}\eqn\FinalAmplitude$$
This formula contains as a special case, an expression
equivalent to an earlier result for
the amplitude with three adjacent negative helicities~[\use\HelicityRecurrence].
I have verified that it agrees numerically with amplitudes computed
via a set of light-cone recurrence relations through $n=10$.

In the unitarity-based method for loop amplitudes~[\use\UnitarityMachinery], 
four-dimensional amplitudes such as the one computed above suffice for
computing one-loop amplitudes in the supersymmetric gauge theories.  For
non-supersymmetric theories, additional amplitudes with
gluon momenta continued to $D=4-2\e$ (or equivalently massive scalars)
are required.

I thank Z.~Bern and L.~Dixon for useful conversations, and the 
the Kavli Institute for Theoretical Physics, where this work was begun,
 for its generous hospitality. 
This research was supported in part by the National Science Foundation 
under Grant No. PHY99-07949.

\listrefs
\bye